\documentclass{sn-jnl}

\usepackage{caption}
\usepackage{booktabs} 
\usepackage{subcaption}
\usepackage{graphicx}%
\usepackage{multirow}%
\usepackage{amsmath,amssymb,amsfonts}%
\usepackage{amsthm}%
\usepackage{mathrsfs}%
\usepackage[title]{appendix}%
\usepackage{xcolor}%
\usepackage{textcomp}%
\usepackage{manyfoot}%
\usepackage{booktabs}%
\usepackage{algorithm}%
\usepackage{algorithmicx}%
\usepackage{algpseudocode}%
\usepackage{listings}%
\usepackage{comment}

\theoremstyle{thmstyleone}%
\theoremstyle{thmstyletwo}%

\theoremstyle{thmstylethree}%

\begin{document}

\title[Measuring Players’ Contribution to Shot Actions in Football]{A Model-Based Restricted Shapley Value to Measure the Players' Contribution to Shot Actions in Football}


\author*[1]{\fnm{Mattia} \sur{Cefis}}\email{mattia.cefis@unibs.it}

\author[2]{\fnm{Rodolfo} \sur{Metulini}}\email{rodolfo.metulini@unibg.it}

\author[1]{\fnm{Maurizio} \sur{Carpita}}\email{maurizio.carpita@unibs.it}

\equalcont{\centering These authors contributed equally to this work. \\ 
\hspace{5mm}
\centering	This version: 16th July, 2026 \\
\centering	Published in\textit{ Computational Statistics} (\url{https://link.springer.com/article/10.1007/s00180-026-01783-x})\\ 
}
\affil[1]{\protect\raggedright 
Department of Economics and Management, University of Brescia, Brescia, Italy}
\affil[2]{\protect\raggedright 
Department of Economics, University of Bergamo, Bergamo, Italy}

\abstract{
This paper proposes a novel framework to assess individual player contributions in football, explicitly accounting for the cooperative nature of shot-ending offensive actions. 
By incorporating team interaction into player evaluation, it also supports economically sustainable decision-making, with practical implications for performance analysis and player scouting. 
Extending the expected Goals (xG) paradigm, we propose the expected Goal Action (xGA), an action-based metric designed to assess the quality of actions through passing networks. 
Furthermore, we adapt cooperative game theory and introduce the Player’s Restricted Shapley Value (PRSV) statistic, a contribution metric based on restricted coalition structures derived from observed passing interactions, where xGA is adopted to compute the cohesion function.
Unlike traditional Shapley approaches, the PRSV one restricts coalitions to tactically admissible player subsets, offering action-specific, interpretable measures of marginal contribution in a cooperative context. 
We apply the framework to 8,421 shot-actions from the Italian League Serie A season 2022/23, and the case studies of AC Milan and SSC Napoli reveal heterogeneity in contributions within teams. 
Combining the PRSV statistic with an individual performance metric highlights the discrepancies between a player's cooperative engagement and goal-conversion ability.}

\maketitle

\section{Introduction}\label{sec:intro}
Football, also known as soccer, is one of the most popular sports worldwide, with the availability of detailed data increasing in recent years. 
The rise of sports analytics has led football to increasingly rely on data-driven approaches in decision-making processes, ranging from athlete performance optimization to scouting, injury analysis, and betting \cite{cefis2022football}.
Nowadays, football teams operate as complex organizations that aim to optimize their resources and maximize success through the use of advanced metrics and analytical tools.
For this reason, football analytics leverages supervised machine learning techniques, exploratory modeling, and advanced data visualization tools to investigate a wide range of topics, including the prediction of event outcomes \cite{leriou2025survival, carpita2015discovering}, the evaluation of performance from a physical perspective \cite{riboli2025training}, and the development of ranking methodologies \cite{carpita2021discovering, pappalardo2019playerank}. 

In this work, we introduce a methodological approach to assess football players’ contributions to shot-ending actions.
From a broader perspective, this type of contribution-based evaluation aligns with the goals of sustainable sports analytics by supporting optimization in resource allocation and enabling more economically efficient recruitment strategies, for example by helping clubs identify players whose collective impact is undervalued by traditional statistics.

Among the specific field of players' performance evaluation, traditional metrics such as goals, assists, and shot-based measures provide a basic understanding but often lack contextualization of each situation. 
Advanced approaches have introduced models such as expected goals (xG), a widely adopted probabilistic framework used by analysts and team staff to quantify shot quality and offensive performance in football \cite{fairchild2018spatial}. Within a model-based framework, shot outcomes are typically estimated using probabilistic classifiers \cite{robberechts2020data, cefis2025accuracy}. Recent studies have applied xG models to investigate several aspects of football performance, including defensive styles and tactical effectiveness in competitive leagues \cite{ruan2022quantifying}.

Modern xG approaches may incorporate different levels of contextual information, ranging from simple shot-location variables to more detailed event-related features. Several studies have focused on improving model precision by introducing contextual variables such as shooter and goalkeeper abilities, the shooter’s visual angle, and other situational characteristics, using both statistical and machine learning approaches \cite{cefis2025accuracy, karim2023kos, cefis2024new}. Another recent work \cite{mead2023expected} showed that the inclusion of additional action-related information, such as possession duration, can further improve xG model performance and practical value.

Despite these developments, most xG approaches still estimate scoring probability mainly from shot-related features, making it difficult to explicitly capture the cooperative build-up process leading to the shot.
In particular, passing structures, player interactions, and the collective organization of offensive sequences are often only partially incorporated, limiting the possibility of evaluating marginal player contributions during attacking actions.

Over the last few years, advanced metrics have been developed to better understand and quantify attacking contributions in football, with a focus on build-up play and positional threat. 
Among these, \textit{xGChain} captures the total expected goals generated by all players involved in the sequence leading up to a shot, assigning the value of the shot equally among all players involved, with each player receiving a share divided by the number of participants \cite{barthelemy2024impact}. 
This metric is useful for evaluating collective contributions and examining tactical build-up, although it may overestimate the influence of players involved in long offensive sequences without explicitly weighting the quality of each contribution.

Differently, \textit{xT} (\textit{Expected Threat}) evaluates the level of danger associated with specific field positions and situations based on spatial progression and ball movement \cite{hassani2025dynamic, chakraborty_xt}. While highly informative for assessing territorial advancement and offensive threat, these approaches provide limited insight into the cooperative interactions among players within the offensive action.

These metrics contribute to exploring the attacking play, but their limitations highlight the need for integrated approaches that consider both spatial and collaborative aspects of football.
To address these limitations, several studies have integrated event and tracking data to account for players’ positions, involvement, and situational context \cite{cefis2024new}. 
However, these approaches remain fragmented and often lack a unified framework capable of capturing the full offensive process and the contextual value of each action. 
Football is inherently a collective game characterized by high variability and unpredictability. The considerable degree of randomness observed in football outcomes, even when using strong predictors such as betting odds \cite{wunderlich2021influence, wunderlich2025using}, implies that even a single pass, deviation, or deflection can change the outcome of a match. This highlights the need for measures capable of capturing both individual performance and the interactions and synergies among players during offensive actions.

In this context, we introduce the expected Goal Action (\textit{xGA}), an action-oriented extension of the \textit{xG} framework that evaluates the full offensive sequence preceding a shot rather than the shot event alone. While maintaining the same probabilistic outcome structure of traditional \textit{xG} models (i.e., goal vs. no goal), \textit{xGA} incorporates action-level features derived from the offensive build-up phase, such as passing structure, player involvement, and spatial progression. Conceptually, \textit{xGA} should not be interpreted as an alternative to the general \textit{xG} paradigm, but rather as an extension in which the unit of analysis shifts from the isolated shot to the entire offensive action leading to the shot. 
\textit{xGA} assigns a scoring probability to the attacking action as a whole by using action-level features, such as the number of players involved and the number of passes, thereby incorporating the contextual and tactical information that characterizes each action.
According to the methodological approach proposed in this work, \textit{xGA} is adopted as the cohesion (or \textit{worth}) function within a metric inspired by cooperative game theory and adapted to the football context, providing a natural framework to quantify each player’s marginal contribution to collective performance. 
In light of this, the distinction between \textit{xG} and \textit{xGA} is particularly relevant, as they are used not only to estimate scoring probability but also to define a cohesion function for measuring players’ marginal contributions within cooperative offensive actions.

Specifically, building on the Shapley value \cite{shapley1953value} and its extensions to restricted cooperation \cite{myerson1977graphs}, we propose the Players' Restricted Shapley Value (PRSV), a cooperative game–theoretic metric for football. 
Its underlying logic aligns well with the nature of this sport, where the outcome of an action depends on the coordinated participation of multiple players.
The Shapley value has been occasionally applied in team sports such as basketball \cite{metulini2023measuring} and football \cite{auer2015evaluation,hiller2015importance}.
In these football-related studies, the worth of a coalition is defined solely by match outcomes, making the resulting Shapley values inherently game-specific. Moreover, the analysis is limited to full team coalitions, overlooking passing structures and the fact that not all players contribute to every action.
Recent work \cite{metulini2024euro} extends this line of research advances this line of research by estimating the worth of each coalition using a model-based approach grounded in the expected-goals framework. 
In the present work, we extend the previous approach by forming player coalitions from the passing network of individual shot actions, thereby assigning payoffs only to players who actively contribute to the development of the play.
This perspective is consistent with action-based approaches proposed in the literature, such as the VAEP framework \cite{decroos2019actions}, which evaluates the value of player actions throughout offensive sequences.

In terms of player evaluation, the proposed framework can be classified as a bottom-up approach according to the distinction introduced by Hvattum \cite{hvattum2019comprehensive}, since player ratings are derived from observed offensive actions and event-level contributions. 
However, unlike traditional event-based bottom-up metrics, our approach explicitly incorporates cooperative interactions and synergies among players based on passing-network coalitions.

Unlike traditional xG-based metrics for individual performance, which primarily reflect a player’s finishing ability (e.g., through xG overperformance), PRSV captures a player’s marginal contribution to the creation of shooting opportunities associated with high scoring probability.
When used alongside standard individual performance metrics, it enables a more comprehensive characterization of players, combining insights into both collaborative impact and finishing ability. 
The proposed approach is evaluated on an original dataset of 8,421 shot actions from the 2022/23 Italian Serie A season, obtained by integrating multiple data sources.
The PRSV statistic is then applied to players from AC Milan and SSC Napoli to illustrate the practical implementation of the framework.
This paper is organized as follows. Section~\ref{sec:method} introduces the proposed $PRSV$ statistic, while Section~\ref{sec:xgamodel} details the expected goal action model used as the cohesion function. Section~\ref{sec:data} describes the dataset, Section~\ref{sec:results} presents the results for the Italian Serie~A with case studies on AC~Milan and SSC~Napoli, and Section~\ref{sec:concl} concludes the paper.

\section{The Player's Restricted Shapley Value (PRSV) statistic} \label{sec:method}

The Shapley value, introduced by Lloyd Shapley in 1953 \cite{shapley1953value}, is a fundamental solution concept in cooperative game theory that provides a fair allocation of collective payoffs among players who cooperate within a coalition in a play. 
It is based on the idea of marginal contribution: each player's value is determined by the average contribution they make when joining all possible subsets, or coalitions, of players.

Formally, let $N = \{1, 2, \dots, n\}$ denote the set of all the players of the grand coalition with cardinality $|N| = n$. 
A \textit{coalition} is any subset $S \subseteq N$ with cardinality $|S| = s \leq n$, and its associated \textit{worth} (\textit{cohesion} or \textit{characteristic}) \textit{function} is represented by $\upsilon(S)$, where $\upsilon : 2^N \rightarrow \mathbb{R}$. 
The Shapley value provides a fair allocation of the total worth among the players, based on their marginal contributions across all possible coalitions, and for a player $i \in N$ is defined as:
  \begin{equation} \label{eq:shapley}
\phi_i(\upsilon) = \sum_{S \subseteq N \symbol{92} \{i\}} \frac{s! (n - s - 1)!}{n!} [\upsilon(S \cup \{i\}) - \upsilon(S)].
  \end{equation}
\par
The summation extends over all subsets $S$ of $N$ that exclude player $i$, and the term ${s!(n - s - 1)!/n!}$ denotes the probability that, in a uniformly random permutation of the \(n\) players, all members of the coalition \(S\) appear \textit{before} player \(i\). 
It is worth noting that, in a team sport context, permutations of players should not be interpreted temporally but as alternative conceptual orderings of functional contribution within an action. 
In other words, this term can be interpreted as a weight assigned to each coalition, determined by its cardinality, such that the total payoff is distributed among the players according to the sizes of their respective coalitions.
\par
The difference $[\upsilon(S \cup {i}) - \upsilon(S)]$ quantifies the marginal contribution of player $i$ to coalition $S$ to be divided among all players. Intuitively, the Shapley value corresponds to the expected marginal contribution of a player when the order of entry of the players is chosen uniformly at random.
\par
According to the classical Shapley value, the properties of global efficiency, players' symmetry, and additivity are satisfied.
Its axiomatic characterization (see Maschler et al. \cite{maschler2020game}, chapter 17) is guaranteed when i) coalitions of any cardinality are considered admissible within the game and so they may have a non-zero worth, ii) all possible coalitions can be observed.
In this setting, the metric can be easily applied to a play. 
However, in applied contexts, particularly in team sports, these conditions are often not met. 
Consider the case of football, where exactly 11 players are on the court together.
This implies that only coalitions composed of exactly 11 players yield a non-zero worth. 
All other coalitions - those with fewer than 11 players, as well as those exceeding 11 players, whether up to the total number of players involved in a single match (typically 14–16, accounting for up to five substitutions) or across an entire season (at least 20) - are assigned zero worth.
Furthermore, not all feasible coalitions of 11 players are necessarily observed throughout a specific time frame (e.g., a game, a season). 
Coaches may favor certain specific coalitions based on preferring complementary players, role constraints, tactics, and styles of play.
Some coalitions tend to appear more frequently, while others may never be observed at all.
As a result, different players will participate in heterogeneous numbers of actions.
Moreover, only those players directly involved in the passing network of a given shot action make an actual contribution. A single action leading to a shot might involve from 1 to 11 players (generally between 3 and 7). 
It follows that, when referring to the Shapley value in Equation \ref{eq:shapley}, the cardinality $n$ is generally equal to 20 or larger (when considering a team during a whole season), while the worth of all the coalitions of cardinality larger than 11 is necessarily zero. 
In football, the empirical distribution of \textit{observed} coalitions across cardinalities differs from the theoretical distribution of all feasible coalitions

In other words, coalitions are not generated randomly, and only a restricted subset of configurations is tactically admissible or empirically observable.
To support this claim, Table~\ref{eq:distr} reports the distribution of observed coalitions in comparison to that of all possible coalitions, for the case of AC Milan of the Italian Serie A season 2022/23 (see section 4), considering the total of 18 selected players. 
The two distributions differ substantially. 
While, from a theoretical perspective, the number of possible coalitions increases with coalition size, the observed coalitions are more frequent for smaller cardinalities.
These figures demonstrate that the observed coalitions are not a random sample (with respect to coalition cardinality) of the total set of possible coalitions. 
This can be attributed, as previously mentioned, to the fact that coaches, due to technical choices and players' roles, do not field all possible coalitions.
\begin{table*}[htbp!] 
\centering
\caption{Distribution of the number of coalitions made by 18 players by their cardinality ($\#$). All possible coalitions (\texttt{all}) versus observed coalitions (\texttt{obs}) based on single actions' passing network in AC Milan (season 2022/23). The largest cardinality is equal to 10 (and not 11) because the goalkeepers are disregarded.}
\begin{tabular}{c | r r | r r}
\hline
\textbf{$\#$} & \texttt{all} & $\%$ & \texttt{obs} & $\%$ \\
\hline
1  & 18     & 0.01& 13 & 3.02 \\
2  & 153    & 0.08 & 50 &11.60 \\
3  & 816    & 0.41 & 72 & 16.71\\
4  & 3,060   & 1.54 & 57 & 13.23 \\
5  & 8,568   & 4.30 & 64 & 14.85\\
6  & 18,564  & 9.32 & 59 & 13.69\\
7  & 31,824  & 15.98 & 40 & 9.28\\
8  & 43,758  & 21.97 & 47 & 10.90\\
9  & 48,620  & 24.42 & 21 & 4.87\\
10 & 43,758  & 21.97 & 6 &1.86\\
\hline
total & 199,139 & 100\% & 431  & 100\% \\
\hline
\end{tabular}
\label{eq:distr}
\end{table*}

The case in which not all possible coalitions can be observed has been addressed in the literature through the introduction of a class of \textit{modified Shapley values for restricted cooperation}\footnote{Differently, the \textit{Random-Order Shapley value} \cite{shapley1953value}, based on sampling random permutations instead of evaluating all coalitions, allows efficient approximation in large systems via Monte Carlo methods \cite{castro2009polynomial}.
Its use is appropriate in contexts where the full computation of the exact Shapley value is too expensive or not feasible due to a large number of players in the considered game, which is not our case.}. 
In this regard, in the spirit of the seminal work by Myerson \cite{myerson1977graphs} (whereas additional examples can be found in the literature, e.g. \cite{derks1993shapley}), in \textit{restricted} formulations of the Shapley value, the unobserved coalitions are simply disregarded. 

We introduce a \textit{restricted} Shapley value, discuss its estimation, and develop a bootstrap-based inferential framework suitable for player comparison and dominance testing.
Using a specific formulation for restricted (or conditional) coalitions, we consider the following equation for player $i$:

\begin{equation} \label{eq:myerson_restr}
\phi_i^{\mathcal{R}}(\upsilon) = \sum_{C \in \mathcal{C}_i} w_i(C)[\upsilon(C\cup \{i\}) - \upsilon(C)],
\end{equation}
where \(\mathcal{R}\) stays for \textit{restricted}, and $\mathcal{C}_i \subseteq 2^{N \setminus \{i\}}$ denotes the set of all the coalitions $C= [(S\cup \{i\}) \backslash \{i\}]$ (some observed, $S$, and some unobserved, $S^\ast$) with cardinality $c = |C|$ \textit{compatible} with the observed coalitions ($S\cup \{i\}$), i.e. the observed actions with player $i$ determined by formation constraints, positions, and coach's tactical decisions.

In Equation \ref{eq:myerson_restr} the Shapley weights are normalized over the restricted support, and admit a probabilistic interpretation, as permutation probabilities conditional on the event that the coalitions of players preceding $i$ are compatible:
\begin{equation}
w_i(C)
=
\frac{\frac{c!(n-c-1)!}{n!}}
{\sum_{C \in \mathcal{C}_i} \frac{c!(n-c-1)!}{n!}}.
\label{eq:normalized_weights}
\end{equation}

In summary, the resulting Shapley value in Equation \ref{eq:myerson_restr} is conditional on the observed situation and should be interpreted as a \emph{context-dependent contribution measure}.
In this case, while global efficiency no longer holds in general, symmetry and marginality are preserved within the restricted support, making this conditional formulation well-suited for this environment, characterized by strategic selection and sparse coalition support.

While restricted Shapley values provide a meaningful decomposition of the coalition outcome, their
raw magnitudes are not directly comparable across players due to heterogeneity in participation and
different coalition supports $\mathcal{C}$. 
Moreover, as explained in the next Section, the worth function $\upsilon$ is \textit{model-based}, that is, it is estimated as $\hat{\upsilon}$ using a statistical model and a sample of observed actions. 
Consequently, the restricted Shapley value in Equation~\ref{eq:myerson_restr} is estimated as $\hat{\phi}_i^{\mathcal{R}} = \phi_i^{\mathcal{R}}(\hat{\upsilon})$, with the \textit{standard error} of this estimator obtained via B bootstrap replications, according to the following formula:

\begin{equation}
\mathrm{SE}(\hat{\phi}_i^{\mathcal{R}})
=
\sqrt{
\frac{1}{B-1}
\sum_{b=1}^B
(\hat{\phi}_i^{\mathcal{R}(b)} - \bar{\phi}_i^{\mathcal{R}})^2},
\label{eq:se_phi}
\end{equation}
with the bootstrap average:
\begin{equation}
\bar{\phi}_i^{\mathcal{R}} = \frac{1}{B} \sum_{b=1}^B \hat{\phi}_i^{\mathcal{R}(b)}.
\label{eq:se_phi_avg}
\end{equation}

The use of bootstrap standard errors is important because the restricted Shapley value $\hat{\phi}_i^R$ depends on the estimated worth function $\hat{\upsilon}$, obtained from the xGA model and a finite sample of observed coalitions. 
Consequently, the estimated contribution is affected by sampling variability arising from model estimation. 
Moreover, players differ substantially in the number and type of actions in which they are involved, leading to heterogeneous estimation uncertainty. Bootstrap resampling therefore provides a flexible way to approximate the sampling distribution of $\hat{\phi}_i^R$ and to quantify the variability of the estimated marginal contributions.
While the restricted Shapley value provides a meaningful decomposition of the collective offensive outcome, its raw magnitude may be difficult to compare across players due to heterogeneous estimation uncertainty. 
Two players may exhibit similar marginal contributions while being characterized by very different levels of variability. 
For this reason, we introduce a studentized measure that evaluates each player's contribution relative to its estimation variability, leading to the definition of the Players' Restricted Shapley Value (PRSV) statistic:

\begin{equation}
PRSV_i
=
\frac{\hat{\phi}_i^{\mathcal{R}}}{\mathrm{SE}(\hat{\phi}_i^{\mathcal{R}})}.
\label{eq:zscore}
\end{equation}

The numerator captures the estimated marginal contribution generated by the player within the restricted coalition structure, while the denominator reflects the uncertainty associated with this estimate.

In addition to accounting for the sampling uncertainty arising from the estimated model and the observed actions, the $PRSV$ statistic is a dimensionless measure that allows for meaningful comparisons of player contributions in units of standard errors.

The $PRSV$ statistic has a useful interpretation as \textit{Signal-to-Noise Ratio}. In fact, the estimator of the conditional Shapley value can be modeled as:

\begin{equation}
\hat{\phi}_i^{\mathcal{R}}
=
\phi_i^{\mathcal{R}} + \varepsilon_i,
\label{eq:signal_error}
\end{equation}
where $\mathbb{E}(\varepsilon_i \mid \mathcal{C}_i)=0$, $\mathrm{Var}(\varepsilon_i \mid \mathcal{C}_i)=\sigma_i^2$ and, under regularity conditions, the bootstrap standard error estimator $\mathrm{SE}(\hat{\phi}_i^{\mathcal{R}})$ is consistent for $\sigma_i$.
Dividing Equation \eqref{eq:signal_error} by $\mathrm{SE}(\hat{\phi}_i^{\mathcal{R}})$ yields:
\begin{equation}
PRSV_i
=
\frac{\phi_i^{\mathcal{R}}}{\sigma_i}
+
\frac{\varepsilon_i}{\sigma_i}
+ o_p(1).
\label{eq:z_signal_noise}
\end{equation}

The term $\phi_i^{\mathcal{R}}/\sigma_i$ represents the \emph{normalized conditional contribution} of player $i$, that is, the magnitude of the contribution measured in units of its intrinsic estimation uncertainty.
The second term, $\varepsilon_i/\sigma_i$, captures the stochastic component of the estimator and has zero mean and unit variance conditional on $\mathcal{C}_i$.
The remainder term $o_p(1)$ arises from replacing the unknown scale parameter $\sigma_i$ with its bootstrap estimator and vanishes asymptotically under consistency of the bootstrap standard error.
Thus the $PRSV$ statistic places players on a common, dimensionless scale that adjusts for heterogeneous uncertainty: large (small) $PRSV$ indicates a contribution that is large (small) relative to its estimation noise.

\section{The action-based xGA model} \label{sec:xgamodel}

 The new action-based \textit{expected Goal Action} (xGA) model is used in this study to estimate the worth or cohesion function $\upsilon$.
This model can be seen as an advancement of the xG that includes aspects such as build-up play, player positioning, and the game situation. 
This enhanced measure thus provides a more comprehensive assessment of the quality and value of shot-related actions.

The xG and the xGA models share the same binary outcome vector $\mathbf{Y}$ of length $H$, with $Y_h = 1$ if the shot action $h$ results in a goal and $Y_h = 0$ otherwise. 
The difference is in the design matrix $\mathbf{X}$ of dimension $H \times K$, whose $K$ columns contain the features describing each action: the present xGA model inherits the $shot$ related features \texttt{X, Y}, \texttt{Shot Angle} from the xG model, and adds the $action$ related features \texttt{first\_pass\_x, first\_pass\_y, passNb, playersNb, avg\_pass\_distance, plPerformanceIndex, h\_a} and \texttt{situation} (see Table \ref{tab:descr_stats} for details). 
As the xG model is $nested$ in the xGA model, only this last one is considered, but the following explanation can be easily extended to the first one.

The xGA model allows to estimate the conditional probability of scoring a goal for a given action $h = 1, 2, \ldots, H$ based on its associated vector of $K$ features $\mathbf{X}_h$:

\begin{equation}\label{eq:xGA_equation}
\widehat{\text{xGA}}_h 
\;=\; \hat{Y}_h
\;=\; \hat{\mathbb{E}}[Y_h \mid \mathbf{X}_h] 
\;=\; \hat{\mathbb{P}}(Y_h = 1 \mid \mathbf{X}_h).
\end{equation}
A given action $h$ here is defined as beginning when a player regains possession of the ball from the opponents and ending with a shot. All actions that do not conclude with a shot are not taken into account.
Following Cefis and Carpita \cite{cefis2025accuracy}, we evaluated different predictive approaches for estimating goal probability, including a traditional statistical model based on Binary Regression with a complementary log-log link function (BR cloglog) \cite{hosmer2013applied} and the XGBoost algorithm \cite{chen2016xgboost}. Since XGBoost achieved superior predictive performance, only the results obtained with this model are reported in the manuscript.

The model was estimated using a training set, while predictive performance was evaluated on a separate test set. Model specification was assessed through the examination of feature importance measures, which quantify the contribution of each predictor to the model’s predictive performance according to how frequently and effectively a feature is used in the construction of the decision trees. These measures provide an intuitive ranking of the most influential factors driving the estimated probability of scoring a goal. 
\par
As stated in the comment to Equation \ref{eq:myerson_restr}, the restricted Shapley value is computed using the set of all the coalitions $C$ (some observed, $S$, and some unobserved, $S^\ast$), \textit{compatible} with the observed coalitions ($S\cup \{i\}$).
Therefore, the \textit{in-sample} estimate in Equation \ref{eq:xGA_equation} is used to estimate the worth function as the sum of the $\widehat{xGA}$ for all the $H_{(S\cup \{i\})}$ observed actions in which player $i$ participated: 

\begin{equation}\label{eq:worth1}
\upsilon(S\cup \{i\}) = \sum_{h=1}^{H_{(S\cup \{i\})}}\widehat{xGA}^{(in)}_{h},
\end{equation}
and for all the $H_{S}$ observed $S = [(S\cup \{i\}) \backslash \{i\}]$:
\begin{equation}\label{eq:worth2}
\upsilon(S) = \sum_{h=1}^{H_{S}}\widehat{xGA}^{(in)}_{h}.
\end{equation}

The \textit{out-of-sample} $\widehat{xGA}$ is used to estimate $\upsilon$ using the $H_{S^*}$ actions that have not been observed but are $compatible$, i.e. unobserved $S^* = [(S\cup \{i\}) \backslash \{i\}]$: 

\begin{equation}\label{eq:worth3}
\upsilon(S^*) = \widehat{xGA}^{(out)}.
\end{equation}

In this equation, we do not have a summation because this worth is estimated on just one fictitious action. Consider the coalition $S^* = [(S\cup \{i\})]$, the value of the features are the same for the corresponding $S^*$, except for the player performance index and the number of players, which are computed on the new set of players.
Note that the superscripts $in$ and $out$ are used to indicate the in-sample and out-of-sample estimates of the xGA, respectively. In other terms, the $quality$ of the estimated worth function in Equations \ref{eq:worth1} and \ref{eq:worth2} rely to the performance of the models on the training set, and the $quality$ of the estimated worth function in Equation \ref{eq:worth3} rely to the performance of the models on the test set.
For this reason, is essential that the models correctly predicts the outcome on the test set.
To check this issue and considering that scoring a goal in football is a rare event (about $10\%$), the approach of Cefis and Carpita \cite{cefis2025accuracy} based on the classification threshold according to the prevalence of the minority class \cite{cavus2022explainable} is used. 
Within this framework, model performance is evaluated on the test set using both traditional classification metrics, namely precision (fraction of labels classified as positive that are true positive), sensitivity or recall (proportion of true positive that are correctly classified), F1 (harmonic mean between precision and recall), specificity (proportion of true negatives that are correctly classified), Area Under the Curve (AUC), and with other metrics suitable for imbalanced data \cite{cavus2022explainable}, namely Mathews Correlation Coefficient (MCC) \cite{chicco2020advantages} and the Brier score (mean squared difference between the observed binary outcome and the predicted probability) \cite{glenn1950verification}.

\section{The actions dataset} \label{sec:data}

To apply and evaluate the proposed method, we constructed an ad hoc dataset that combines event data from all matches of the Italian Serie A season 2022/23 (specifically, all actions leading to shots) with player performance data.
We focus on the Italian Serie A, because the league is considered one of the top football leagues worldwide, ranked 2$^{\text{nd}}$ according to UEFA ranking \textit{uefa.com}. 
During the 2022/23 season, the league consisted of 20 teams, each playing 38 matches, with one home and one away match against every other team.
To create the final structured dataset, we combined data from multiple sources, following the typical Extraction Transformation Loading (ETL) process, commonly used in data engineering and computer science \cite{kimball2008data}. In our study, this phase included: (i) extracting data from multiple sources, (ii) harmonizing and integrating datasets through unique identifiers, (iii) cleaning and preprocessing the data, and (iv) generating derived variables used for the subsequent analyses.
In particular, event and passing network data were obtained via web scraping from \textit{whoscored.com}, a well-known football statistics website. 
Additional context-related event data and expected goals were extracted from \textit{understat.com} using the \textit{R} package \texttt{worldfootballR}. Performance data were obtained from \textit{sofifa.com}, which integrates subjective ratings from over 9,000 scouts, coaches, and fans. 
These ratings were used to construct composite indicators based on 29 Key Performance Indicators (KPIs) for movement players and 31 for goalkeepers. The indicators were derived using a hierarchical latent structure estimated through a PLS-SEM approach, as defined and validated in \cite{cefis2024higher}, with values updated up to the date of each match \cite{biecek2021explanatory}. Field coordinates, based on \textit{understat.com} and \textit{whoscored.com} rules, employed width as the $y$-axis and length as the $x$-axis, both expressed as percentages of the pitch, considering offensive actions standardized from left to right.
The dataset consists of 8,421 shot-actions with the following features:
\begin{itemize}
\item Event features from \textit{Whoscored}: the number of passes (\texttt{passNb}) and the number of players involved in each action (\texttt{playersNb}); then, the average pass distance \texttt{avg\_pass\_distance} and the first pass action on the pitch (\texttt{first\_pass\_x} and \texttt{first\_pass\_y}), the players' id and role which are involved in the actions and the match-game timing of each action.

\item Event features from \textit{Understat}: some context features, such as the match name and date, the shooter, the assist-man, and the team names involved in the actions; the shot starting point on the pitch (used for evaluating the proximity to goal, by coordinates \texttt{X} and \texttt{Y}), the \texttt{Shot Angle}, the action \texttt{situation} (e.g. open play, free kick, penalty, etc.), the shot type (e.g. right, left, head or other), the final binary outcome (\texttt{Outcome}) of each shot (goal or not), and whether the team plays at home or away (\texttt{h\_a}).

\item Performance feature from \textit{Sofifa}: for each action, \texttt{plPerformanceIndex} is the variable representing
the mean of the offensive performance index of the teammates involved in the action. Note that for the actions with only one player involved, this feature corresponds to that player’s offensive index.
\end{itemize}

In Table \ref{tab:descr_stats} we provide a summary about the main features of the dataset. Specifically, Table \ref{tab:continuous_var} shows
details on the continuous variables, while Table \ref{tab:categoric_var} focuses on the categorical variables. Note that features in bold have been computed during the ETL phase. 
A more detailed description of this dataset is given in Cefis et al. \cite{cefis2025new}.

\begin{table*}[htbp!]
\centering
\caption{Overview of features with description, source, type, and descriptive statistics for the dataset of 8,421 shot actions from the Italian Serie A season 2022/23. Features in bold have been computed during the
ETL phase.}
\label{tab:descr_stats}
\begin{subtable}{\linewidth}
\centering
\caption{Continuous features.\label{tab:continuous_var}}

\resizebox{\linewidth}{!}{%
\begin{tabular}{|l|p{3.0cm}|l|l|cccc|}
\hline
\textbf{Feature} & \textbf{Description} & \textbf{Source} & \textbf{Type} &
\textbf{Mean} & \textbf{St. Dev.} & \textbf{Skewness} & \textbf{Kurtosis} \\ \hline
X & X shot starting coordinate & Understat & Shot & 85.31 & 7.39 & -0.41 & -0.82 \\ \hline
Y & Y shot starting coordinate & Understat & Shot & 50.84 & 12.42 & 0.00 & -0.50 \\ \hline
Shot Angle & Shot Angle (degree) & Understat & Shot & 33.77 & 20.98 & 0.20 & -0.94 \\ \hline
\textbf{first\_pass\_x} & X coordinate of the first action pass & WhoScored & Action & 51.15 & 27.12 & 0.10 & -0.94 \\ \hline
\textbf{first\_pass\_y} & Y coordinate of the first action pass & WhoScored & Action & 50.79 & 29.34 & -0.04 & -0.93 \\ \hline
\textbf{passNb} & Number of passes & WhoScored & Action & 6.47 & 5.85 & 2.00 & 6.44 \\ \hline
\textbf{playersNb} & Number of players & WhoScored & Action & 4.87 & 2.52 & 0.39 & -0.79 \\ \hline
\textbf{avg\_pass\_distance} & Mean passes distance in the action & WhoScored & Action & 27.28 & 10.60 & 1.39 & 2.99 \\ \hline
\textbf{plPerformanceIndex} & Mean players-action offensive performance index & Sofifa & Action & 84.81 & 9.39 & -1.36 & 4.17 \\ \hline
\end{tabular}
}
\end{subtable}

\vspace{0.6cm}

\begin{subtable}{\linewidth}
\centering
\caption{Categorical features.\label{tab:categoric_var}}

\resizebox{\linewidth}{!}{%
\begin{tabular}{|l|p{5.1cm}|l|l|cccc|}
\hline
\textbf{Feature} & \textbf{Description} & \textbf{Source} & \textbf{Type} &
\textbf{freq(1)} & \textbf{freq(2)} & \textbf{freq(3)} & \textbf{freq(4)} \\ \hline
Outcome & Goal(1) No Goal(2) & Understat & Outcome & 0.10 & 0.90 & - & - \\ \hline
\textbf{h\_a} & Home(1) Away(2) & Understat & Action & 0.54 & 0.46 & - & - \\ \hline
\textbf{situation} & Open Play(1) Free Kick(2) Penalty(3) Others(4) & Understat & Action & 0.73 & 0.09 & 0.01 & 0.17 \\ \hline
\end{tabular}
}
\end{subtable}

\end{table*}


\section{Empirical analysis} \label{sec:results}

\subsection{The xGA model results}\label{sec:xga}
Preliminarily, the model used for calculating xGA probabilities was subjected to a detailed evaluation process. 
First, the absence of multicollinearity among the regressors defined in Table \ref{tab:descr_stats} was confirmed thanks to the classical Variance Inflation Factor (VIF) index, which turned out to be lower than 4 for all the features.
To account for stability and robustness in the results, model performance was evaluated using a bootstrap procedure with B = 1,000 replications. 
At each iteration, the bootstrap sample was used for training, while the remaining observations (approximately the $36.8\%$ of data) served as the test set for computing performance metrics. 
Figure \ref{fig:var_importance} reports the 90\% bootstrap confidence intervals of feature importance, ordered in decreasing order by their values.
Proximity to goal (variable \texttt{X}) clearly emerges as the most influential predictor, followed by \texttt{shot\_angle}, both core variables of the baseline xG model. 
Additional action-contextual variables, such as \texttt{actionSituationCategory}, \texttt{avg\_pass\_distance}, \texttt{plPerformanceIndex}, and the origin of the action (coordinates \texttt{first\_pass\_x} and \texttt{first\_pass\_y} on the pitch), exhibit moderate to low importance. Conversely, \texttt{h\_a} and \texttt{plaersNb} appear to have negligible relevance, with importance values close to zero. 

\begin{figure}[htbp!]
    \centering
    \centering
    \includegraphics[width=\textwidth]{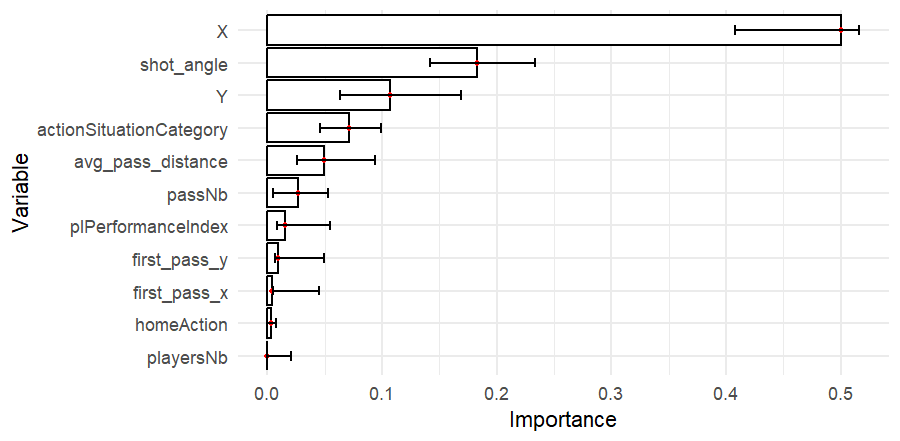}
        
        \label{fig:var_importance}
    \caption{Features importance: original sample estimates (horizontal bars) and 90\% bootstrap percentiles intervals (B = 1,000 bootstrap replications).}
    \label{fig:var_importance}
\end{figure}

The results reported in Table \ref{tab:perf_metrics} summarize the bootstrap mean estimates of the performance metrics computed on the test set, together with their corresponding bootstrap standard errors. Particularly, the xGA model shows satisfactory predictive performance, achieving a good balance between sensitivity and specificity, as well as a relatively high Area Under the Curve ($AUC = 0.79$), indicating strong discriminative ability. Moreover, the positive Matthews Correlation Coefficient ($MCC = 0.24$) suggests a meaningful predictive association even in the presence of class imbalance, while the low Brier score indicates adequate calibration of the predicted probabilities. The relatively small bootstrap standard errors across all metrics further indicate good stability and robustness of the model performance estimates over the resampling procedure.

\begin{table*}[htbp!]
\small
\centering
\caption{Performance estimates (Est.) and Standard Error Bootstrap (SEB) of the xGA model on the test set. 
B=1,000 bootstrap replications.}
\label{tab:perf_metrics}

\begin{tabular}{|l|c|c|}
\hline
\multirow{2}{*}{\textbf{Metric}} 
& \multicolumn{2}{c|}{\textbf{xGA model}} \\
& Est. & SEB \\
\hline
Sensitivity & 0.79 & 0.033 \\
Specificity & 0.62 & 0.026 \\
F1          & 0.28 & 0.012 \\
Precision   & 0.17 & 0.009 \\
MCC         & 0.24 & 0.014 \\
AUC         & 0.79 & 0.012 \\
Brier       & 0.07 & 0.003 \\
\hline
\end{tabular}

\end{table*}

\subsection{The PRSV statistic results for AC Milan and SSC Napoli}\label{sec:case_studies}

In this sub-Section, we present the results related to the application of the $PRSV$ statistic defined in Equation \ref{eq:zscore} to two Serie A teams during season 2022/23.
We selected AC Milan, a club recognized worldwide and the most successful Italian team in international competitions, and SSC Napoli, the league champion of that season.
Victor Osimhen, the forward of SSC Napoli, scored the highest number of goals in the league (26), while AC Milan finished fourth that season, registering 64 goals scored and 43 conceded. Both teams exhibited a common feature: their offensive production was distributed across a high number of players, with 15 players per team involved in at least 60 shooting actions over the season\footnote{The threshold of 60 shots corresponds to the 25th percentile of the distribution of players’ shot actions.}. 

Some restriction has been imposed to the sample of actions and players considered in this analysis. 
First, penalty kicks are excluded, as they represent a very specific and isolated situation that does not reflect the cooperative build-up phase of offensive actions (just 1.1\% of shots in our sample is represented by penalty kicks). Among players, goalkeepers were excluded due to the distinct nature of their role in relation to the creation of shot actions.
Moreover, to ensure robustness in the evaluation of players’ contributions, we restricted the sample to just players who were involved in at least 60 offensive actions during the season (that corresponds to 15 players for both teams). 
This filtering criterion guarantees that the computed $PRSV$ values are not biased by the inclusion of players with a very limited participation in the passing network.
During the observed season, AC Milan produced a high number of shot-actions (505), generated by 414 distinct observed coalitions, while additional 1,800 out-of-sample coalitions were constructed, allowing for a detailed assessment of both observed and unobserved cooperative structures; concerning SSC Napoli, it produced an even higher number of shot-actions (519), generated by 410 distinct observed coalitions, and 1,834 are the out-of-sample coalitions.

The $PRSV$ statistic has been computed for all 30 considered players (15 from AC Milan and 15 from SSC Napoli) using the bootstrap with B = 1,000 replications, aimed to estimate the standard error in Equations \ref{eq:se_phi} and \ref{eq:se_phi_avg}.
Table~\ref{tab:milan_players_summary} reports, for each player, the number of shot actions in which they were involved, their role, and their $PRSV$ statistic.
It is worth noting that each player is consistently assigned to the same role $j$ across all considered actions. 
This is because a player's role is determined based on the information obtained from \textit{Sofifa} (as detailed in Section \ref{sec:data}), rather than the specific role occupied in a given action.

The results reveal a marked heterogeneity in players’ contributions to shot quality generation, both within and across teams. At the top of the overall ranking, Victor Osimhen clearly stands out, exhibiting the highest $PRSV$ statistic among all players. This finding highlights his central role in SSC Napoli’s attacking structures, consistent with a system strongly oriented around his ability to finalize and elevate the quality of offensive actions. Among AC Milan players, Olivier Giroud and Rafael Leão display comparably high $PRSV$ statistic, confirming their importance in the attacking phase. 
\begin{table}[htbp!]
\centering
\caption{Role (For: Forward; Mid: Midfielder; Def: Defender), number of actions, and $PRSV$ statistic for the players of AC Milan and SSC Napoli (B = 1,000 bootstrap replications).}
\label{tab:milan_players_summary}
{%
\begin{tabular}{l l c c |l l c c}
\hline
\multicolumn{4}{c|}{\textbf{AC Milan}} & \multicolumn{4}{c}{\textbf{SSC Napoli}} \\
\textbf{Player} & \textbf{Role} & \textbf{Actions} & 
\textit{\textbf{PRSV}} 
& 
\textbf{Player} & \textbf{Role} & \textbf{Actions} & 
\textit{\textbf{PRSV}} 
 \\
\hline
O. Giroud & For & 142 & 3.81 & V. Osimhen & For & 192 & 4.84 \\
R. Leão & For & 201 & 3.44 & K. Kvaratskhelia & For & 216 & 1.53 \\
F. Tomori & Def & 151 & 2.95 & F. Anguissa & Mid & 243 & 1.48 \\
C. De Ketelaere & Mid & 84 & 2.70 & M. Kim & Def & 201 & 1.21 \\
B. Díaz & Mid & 147 & 2.48 & M. Politano & For & 111 & 0.28 \\
S. Tonali & Mid & 202 & 0.14 & T. Ndomelè & Mid & 61 & 0.27 \\
T. Hernández & Def & 202 & -0.29 & E. Elmas & Mid & 104 & 0.24 \\
A. Rebic & For & 66 & -0.39 & M. Olivera & Def & 109 & 0.15 \\
R. Krunic & Mid & 90 & -0.83 & H. Lozano & For & 117 & -0.53 \\
M. Thiaw & Def & 77 & -1.02 & M. Rui Silva Duarte & Def & 168 & -0.73 \\
P. Kalulu & Def & 142 & -1.27 & P. Zielinski & Mid & 235 & -1.36 \\
A. Saelemaekers & For & 116 & -1.28 & S. Lobotka & Mid & 236 & -2.11 \\
D. Calabria & Def & 117 & -1.83 & G. Di Lorenzo & Def & 239 & -2.25 \\
I. Bennacer & Mid & 171 & -3.19 & J. Jesus & Def & 63 & -2.70 \\
J. Messias & For & 97 & -4.39 & A. Rrahmani & Def & 155 & -3.62 \\

\hline
\end{tabular}
}
\end{table}
Players such as Charles De Ketelaere and Brahim Díaz rank among the most influential AC Milan midfielders. On the SSC Napoli side, Frank Anguissa and Khvicha Kvaratskhelia show positive $PRSV$ values, indicating their importance in supporting and sustaining high-quality attacking sequences.
Defensive players also exhibit a non-negligible $PRSV$ statistic, particularly in AC Milan, where Fikayo Tomori and Theo Hernández rank relatively high. This suggests a significant involvement of defenders in cooperative structures leading to shot generation, likely through ball recovery, progression, and support in advanced phases. In SSC Napoli, Min-jae Kim emerges as the most influential defender, reflecting his role in stabilizing possession and enabling structured build-up phases.
Moving down the table, several players from both teams exhibit $PRSV$ values close to zero or negative. 
Players with negative $PRSV$ values, such as Junior Messias and Ismaël Bennacer for AC Milan, and Stanislav Lobotka and Amir Rrahmani for SSC Napoli, are those associated with coalitions that generate below-average shot quality.

We compared the $PRSV$ statistic with the existing xGChain metric from Understat (\url{understat.com}) described in Section \ref{sec:intro}, normalized per number of actions. 
This comparison revealed moderate-to-low Spearman rank correlations ($\rho_S = 0.38$, for AC Milan and $\rho_S = 0.42$, for SSC Napoli). These results could be expected, as xGChain mainly reflects involvement in offensive sequences leading to a shot, without explicitly accounting for action features. In contrast, $PRSV$ evaluates players through a cooperative and action-specific framework based on estimated marginal contributions. Therefore, the moderate correlations suggest that the two metrics capture related but different dimensions of offensive performance.

These results confirm the ability of the proposed $PRSV$ statistic to disentangle not only the magnitude but also the stability of individual contributions within complex cooperative attacking structures, while also highlighting structural differences between AC Milan’s and SSC Napoli’s offensive organization.
\par
\begin{figure}[htbp!]
\centering
\begin{subfigure}[b]{\linewidth}
    \centering
    \includegraphics[width=\linewidth]{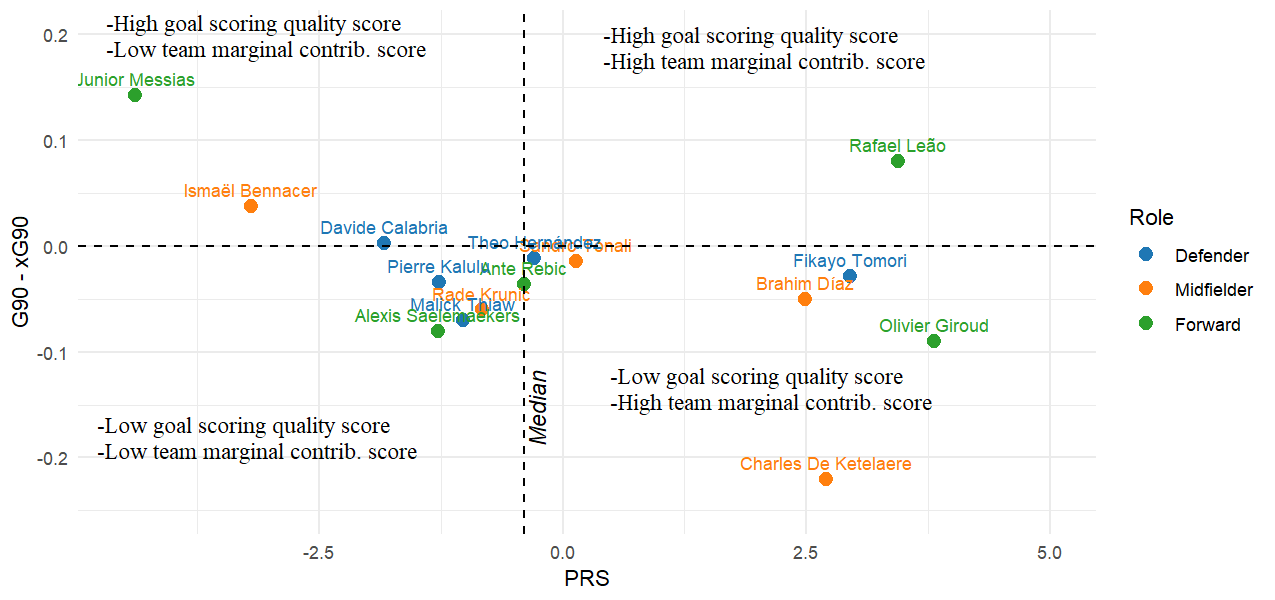}
    \caption{AC Milan players, season 2022/23}
    \label{fig:scatterplot_Milan}
\end{subfigure}
\vspace{0.4cm}
\begin{subfigure}[b]{\linewidth}
    \centering
\includegraphics[width=\linewidth]{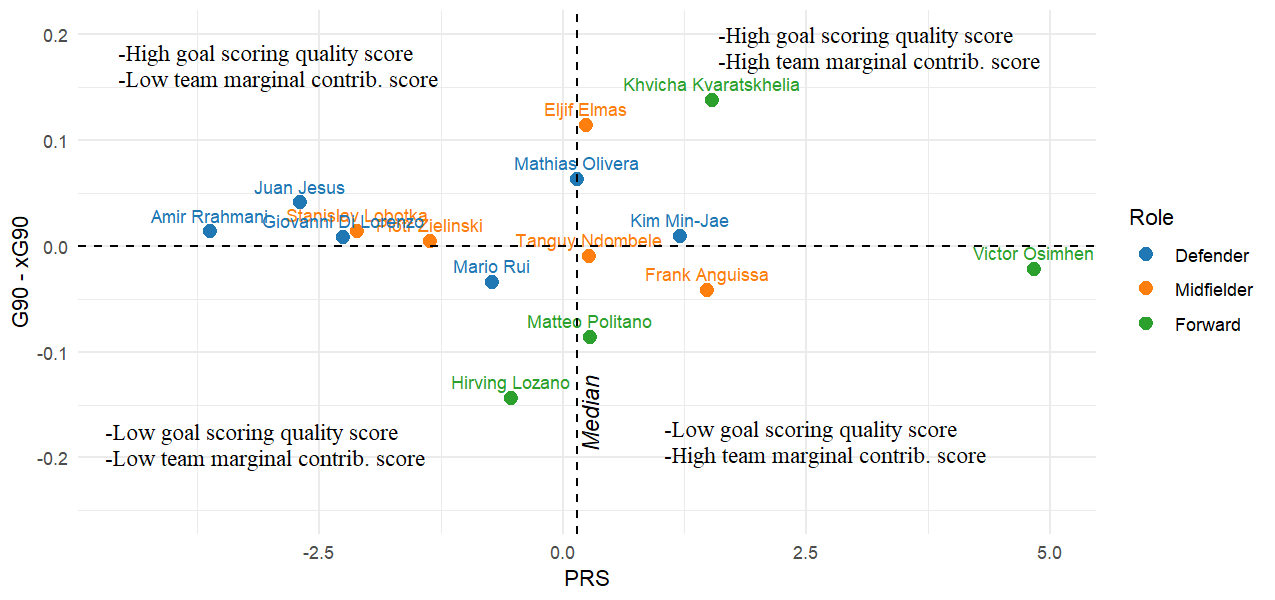}
    \caption{SSC Napoli players, season 2022/23}
    \label{fig:scatterplot_Napoli}
\end{subfigure}
\caption{Scatterplot of the $PRSV$ statistics (B = 1,000 bootstrap replications) (x-axis) and the score efficiency $(G90 - xG90)$ (y-axis).}
\label{fig:scatterplot_}
\end{figure}

The $PRSV$ statistic can be combined with other xG-based metrics to provide useful insights into players’ characteristics, thereby supporting coaching staff and scouting departments in the decision-making process.
Specifically, a more traditional measure of overperformance, namely the difference between goals scored (G) and the traditional expected goals (xG) per 90 minutes \cite{cefis2025accuracy} is considered. 
The resulting scatterplot, where $G90 - xG90$ is reported on the $Y$-axis and the $PRSV$ statistic on the $X$-axis, is presented in Figure \ref{fig:scatterplot_}.
This visualization allows us to jointly assess whether a player's cooperative involvement in offensive actions (through $PRSV$) aligns with, or diverges from, their score efficiency (i.e.,  $G90 - xG90$), which measures individual finishing ability. 
For clarity, we added a vertical dotted line corresponding to the median of the $PRSV$ statistics (with the horizontal dotted line set at 0), and we annotated each quadrant with explanatory labels to better highlight the different combinations of contribution to shot action and individual finishing ability. 
The plot highlights several remarkable patterns. 
Focusing first on AC Milan (Fig. \ref{fig:scatterplot_Milan}), Leão clearly stands out in the top-right quadrant, combining strong goal-scoring ability with a high marginal contribution to the team. This positioning confirms his status as a top player who successfully integrates individual finishing quality with cooperative involvement, as reflected by his 15 goals and 8 assists during the season.
The top-left quadrant includes players such as Messias and Bennacer, who display positive finishing efficiency despite very low $PRSV$ statistic, suggesting limited involvement in the collective build-up phase. Conversely, the bottom-right quadrant comprises players such as De Ketelaere, Brahim Díaz, Tomori, and Giroud, characterized by high $PRSV$ values but negative  $G90 - xG90$. These players contribute substantially to the creation and development of offensive actions, despite underperforming in terms of individual finishing efficiency.
Players located in the bottom-left quadrant exhibit both low marginal contribution and weak finishing efficiency, indicating a limited offensive impact along both dimensions, with Saelemaekers representing the most evident case. In addition, several players cluster around the origin of the axes, reflecting intermediate performance levels, with neither a pronounced cooperative involvement nor a clear advantage in individual finishing efficiency.

Considering SSC Napoli (Fig. \ref{fig:scatterplot_Napoli}), a different configuration emerges. Players are more widely dispersed across the plane, and the only clearly identifiable cluster is located in the top-left quadrant, which includes five players. Kvaratskhelia confirms his profile as a top player by appearing in the top-right quadrant, alongside Elmas, Kim, and Olivera. Osimhen records the highest contribution score, emphasizing his central role in SSC Napoli’s attacking dynamics (in line performance in terms of goal scoring quality), a pattern similarly observed for Anguissa. Politano and Lozano show a median level of contribution to the team’s play, but a negative goal-scoring efficiency. Finally, Mario Rui is the only player clearly positioned in the bottom-left quadrant, indicating a limited offensive impact along both dimensions, although his position remains close to the origin of the axes.  

These patterns may provide useful insights for coaches, including tactical adjustments and individual player evaluation.

\section{Concluding remarks} \label{sec:concl}

In this paper, we introduced a new framework for evaluating individual player contributions in football that accounts for the cooperative nature of offensive actions. By offering valuable insights into players’ collaborative impact, the proposed approach supports data-driven and economically sustainable decision-making for players' scouting and resource allocation, aligning advanced analytics with wider goals of efficiency and sustainability in sports management. 

First, we proposed the \textit{expected Goal Action} (xGA), a novel metric that extends the classic expected goals (xG) model by providing a contextualized measure of shot quality that incorporates information on build-up play, passing structure, and team context.  
This metric was then embedded within a cooperative game-theoretic framework through the definition of a \textit{Player’s Restricted Shapley Value} (PRSV) statistic, designed to allocate the collective worth of an offensive action among the players actively involved in its development.

From a methodological perspective, the main contribution of this work lay in adapting the Shapley value to a setting characterized by restricted and non-random coalition structures.  
By leveraging observed passing networks, we defined coalitions that are both tactically admissible and empirically relevant.  
The resulting $PRSV$ statistic provides an action-specific and context-dependent measure of marginal contribution, allowing for a more granular and realistic attribution of worth in a highly cooperative environment.

Empirically, the application to the Italian Serie A 2022/23 season demonstrated the practical relevance of the proposed approach.  
The empirical analysis showed that the proposed xGA model provides a robust and well-calibrated probabilistic representation of offensive actions, while incorporating contextual information related to build-up play and cooperative structures.  
When used as the cohesion function within the $PRSV$ framework, xGA enables the identification of substantial heterogeneity in players’ contributions, both within and across teams.  
The selected case studies on AC Milan and SSC Napoli highlighted how the proposed metric captures distinct tactical roles, rewarding players who consistently contribute to the creation of high-quality offensive actions through build-up involvement and coordination.

Beyond ranking players by their marginal contribution, the use of bootstrap-based inference allowed us to assess the stability and uncertainty of $PRSV$ statistic estimates, providing additional insight into the robustness of player evaluations. This aspect may be particularly relevant in applied settings, such as scouting, tactical analysis, and performance monitoring, where decision-making often requires an understanding not only of expected contribution levels but also of their uncertainty.

Several limitations of the present study point to promising directions for future research. First, the analysis is restricted to shot-ending actions and does not explicitly account for possession sequences that do not culminate in a shot, which may nonetheless be tactically valuable. Extending the framework to include possession-based or threat-based cohesion functions could further enrich the evaluation of player contributions. Second, coalitions are defined based on passing networks alone; incorporating tracking data to model off-ball movements and spatial interactions would allow for a more comprehensive representation of cooperation. Looking ahead, several avenues for development are possible: expanding the dataset to cover multiple seasons, additional competitions, and a larger number of lineups and shot actions would strengthen the robustness and generalizability of the findings, leading to give emphasis also to match-level situations. Finally, linking the Shapley framework to scouting and decision-making contexts could make the measure more actionable, allowing coaches and analysts to identify players who excel in cooperative involvement even if their goal conversion metrics are modest, or, conversely, to detect players whose efficiency in front of goal masks limited contribution to team build-up.

In summary, the proposed $PRSV$ statistic, adapted to the football context, represents an original and useful contribution to performance evaluation. It complements existing metrics and opens new possibilities for understanding the balance between individual quality and collective involvement. Despite a few limitations, the framework provides a solid foundation for future developments and holds promise as a tool for advancing both academic research and practical applications in football analytics.

\section*{Acknowledgments}
The authors thank the participants of the EURO24 conference in Copenhagen, MathSport International 2025 in Luxembourg, and IES 2025 in Brixen.

\bibliographystyle{unsrt}

\begin{thebibliography}{99}
	
	\bibitem{cefis2022football}
	Cefis, M. (2022).
	Football analytics: a bibliometric study about the last decade contributions.
	\textit{Electronic Journal of Applied Statistical Analysis}, 15(1), 232--248.
	
	\bibitem{leriou2025survival}
	Leriou, I., Ntzoufras, I. (2025).
	Survival modeling of goal arrival times in English premier league.
	\textit{Computational Statistics}, 40(4), 2109--2133.
	
	\bibitem{carpita2015discovering}
	Carpita, M., Sandri, M., Simonetto, A., Zuccolotto, P. (2015).
	Discovering the drivers of football match outcomes with data mining.
	\textit{Quality Technology \& Quantitative Management}, 12(4), 561--577.
	
	\bibitem{riboli2025training}
	Riboli, A., Nardi, F., Osti, M., Cefis, M., Tesoro, G., Mazzoni, S. (2025).
	Training load, official match locomotor demand, and their association in top-class soccer players during a full competitive season.
	\textit{The Journal of Strength \& Conditioning Research}, 39(2), 249--259.
	
	\bibitem{carpita2021discovering}
	Carpita, M., Golia, S. (2021).
	Discovering associations between players' performance indicators and matches' results in the European Soccer Leagues.
	\textit{Journal of Applied Statistics}, 48(9), 1696--1711.
	
	\bibitem{pappalardo2019playerank}
	Pappalardo, L., Cintia, P., Ferragina, P., Massucco, E., Pedreschi, D., Giannotti, F. (2019).
	PlayeRank: data-driven performance evaluation and player ranking in soccer via a machine learning approach.
	\textit{ACM Transactions on Intelligent Systems and Technology (TIST)}, 10(5), 1--27.
	
	\bibitem{fairchild2018spatial}
	Fairchild, A., Pelechrinis, K., Kokkodis, M. (2018).
	Spatial analysis of shots in MLS: a model for expected goals and fractal dimensionality.
	\textit{Journal of Sports Analytics}, 4(3), 165--174.
	
	\bibitem{robberechts2020data}
	Robberechts, P., Davis, J. (2020).
	How Data Availability Affects the Ability to Learn Good xG Models.
	In \textit{Machine Learning and Data Mining for Sports Analytics}, pp.~17--27.
	Springer.
	
	\bibitem{cefis2025accuracy}
	Cefis, M., Carpita, M. (2025).
	Accuracy and explainability of statistical and machine learning xG models in football.
	\textit{Statistics}, 59(2), 426--445.
	
	\bibitem{ruan2022quantifying}
	Ruan, L., Ge, H., Shen, Y., Pu, Z., Zong, S., Cui, Y. (2022).
	Quantifying the effectiveness of defensive playing styles in the Chinese Football Super League.
	\textit{Frontiers in Psychology}, 1--10.
	
	\bibitem{karim2023kos}
	Karim, H., Lotfi, M. (2023).
	The Kos Angle, an optimizing parameter for football expected goals (xG) models.
	\textit{International Journal of Computer Science in Sport}, 22(2), 49--61.
	
	\bibitem{cefis2024new}
	Cefis, M., Carpita, M. (2024).
	A new xG model for football analytics.
	\textit{Journal of the Operational Research Society}, 1--13.

\bibitem{mead2023expected}
Mead, J., O'Hare, A., \& McMenemy, P. (2023).
Expected goals in football: Improving model performance and demonstrating value.
\textit{PLOS ONE}, \textbf{18}(4), e0282295.


	\bibitem{barthelemy2024impact}
	Barthelemy, B., Rav{\'e}, G., Govindasamy, K., Ali, A., Del Coso, J., Demeaux, J., Bideau, B., Zouha, H. (2024).
	Impact of technical-tactical and physical performance on the match outcome in professional soccer: A case study.
	\textit{Journal of Human Kinetics}, 94, 203.
	
	\bibitem{hassani2025dynamic}
	Hassani, K., Ramdani, M., Lotfi, M. (2025).
	Dynamic Expected Threat (DxT) Model: Addressing the Deficit of Realism in Football Action Evaluation.
	\textit{Applied Sciences}, 15(8), 4151.
	
	\bibitem{chakraborty_xt}
	Chakraborty, K. (n.d.).
	Expected Threat (xT): The Value of a Soccer Possession.
	Available at: \texttt{https://karun.in/blog/expected-threat.html}. Accessed: 2025-07-30.

\bibitem{wunderlich2021influence}
Wunderlich, F., Seck, A., \& Memmert, D. (2021).
The influence of randomness on goals in football decreases over time: An empirical analysis of randomness involved in goal scoring in the English Premier League.
\textit{Journal of Sports Sciences}, \textbf{39}(20), 2322--2337.

\bibitem{wunderlich2025using}
Wunderlich, F. (2025).
Using the wisdom of crowds in sports: How performance analysis in football can benefit from the information enclosed in betting odds.
\textit{International Journal of Performance Analysis in Sport}, \textbf{25}(4), 687--706.


	\bibitem{shapley1953value}
	Shapley, L.~S. (1953).
	A value for $n$-person games.
	In H.~W. Kuhn and A.~W. Tucker (Eds.), \textit{Contributions to the Theory of Games}, pp.~307--317.
	Princeton University Press.
	
	\bibitem{myerson1977graphs}
	Myerson, R.~B. (1977).
	Graphs and cooperation in games.
	\textit{Mathematics of Operations Research}, 2(3), 225--229.
	
	\bibitem{metulini2023measuring}
	Metulini, R., Gnecco, G. (2023).
	Measuring players' importance in basketball using the generalized Shapley value.
	\textit{Annals of Operations Research}, 325(1), 441--465.
	
	\bibitem{auer2015evaluation}
	Auer, B.~R., Hiller, T. (2015).
	On the evaluation of soccer players: a comparison of a new game-theoretical approach to classic performance measures.
	\textit{Applied Economics Letters}, 22(14), 1100--1107.
	
	\bibitem{hiller2015importance}
	Hiller, T. (2015).
	The importance of players in teams of the German Bundesliga in the season 2012/2013: a cooperative game theory approach.
	\textit{Applied Economics Letters}, 22(4), 324--329.
	
	\bibitem{metulini2024euro}
	Metulini, R., Cefis, M. (2024).
	Some novelty on the XG model for Football Analytics.
	In \textit{EURO 2024 Conference Handbook \& Abstracts}, pp.~248--249.

\bibitem{decroos2019actions}
Decroos, T., Bransen, L., Van Haaren, J., \& Davis, J. (2019).
Actions speak louder than goals: Valuing player actions in soccer.
In \textit{Proceedings of the 25th ACM SIGKDD International Conference on Knowledge Discovery \& Data Mining} (pp. 1851--1861).


\bibitem{hvattum2019comprehensive}
Hvattum, L. M. (2019).
A comprehensive review of plus-minus ratings for evaluating individual players in team sports.
\textit{International Journal of Computer Science in Sport}, \textbf{18}(1), 1--23.

	\bibitem{maschler2020game}
	Maschler, M., Zamir, S., Solan, E. (2020).
	\textit{Game Theory}.
	Cambridge University Press.

\bibitem{castro2009polynomial}
Castro, J., G{\'o}mez, D., Tejada, J. (2009).
Polynomial calculation of the Shapley value based on sampling. 
\textit{Computers \& operations research}, 36(5), 1726--1730.
    
	\bibitem{derks1993shapley}
	Derks, J., Peters, H. (1993).
	A Shapley value for games with restricted coalitions.
	\textit{International Journal of Game Theory}, 21(4), 351--360.
	
	
	\bibitem{hosmer2013applied}
	Hosmer, D.~W., Lemeshow, S., Sturdivant, R.~X. (2013).
	\textit{Applied Logistic Regression}.
	John Wiley \& Sons.
	
	\bibitem{chen2016xgboost}
	Chen, T., Guestrin, C. (2016).
	XGBoost: A scalable tree boosting system.
	In \textit{Proceedings of the 22nd ACM SIGKDD}, pp.~785--794.
	
	\bibitem{cavus2022explainable}
	Cavus, M., Biecek, P. (2022).
	Explainable expected goal models for performance analysis in football analytics.
	In \textit{2022 IEEE 9th DSAA}, pp.~1--9.
	
	\bibitem{chicco2020advantages}
	Chicco, D., Jurman, G. (2020).
	The advantages of the Matthews correlation coefficient (MCC).
	\textit{BMC Genomics}, 21, 1--13.
	
	\bibitem{glenn1950verification}
	Brier, G.~W. (1950).
	Verification of forecasts expressed in terms of probability.
	\textit{Monthly Weather Review}, 78(1), 1--3.

\bibitem{kimball2008data}
Kimball, R., Ross, M., Thornthwaite, W., Mundy, J., \& Becker, B. (2008).
\textit{The Data Warehouse Lifecycle Toolkit}.
John Wiley \& Sons.
    
	\bibitem{cefis2024higher}
	Cefis, M., Carpita, M. (2024).
	The higher-order PLS-SEM confirmatory approach for composite indicators of football performance quality.
	\textit{Computational Statistics}, 39(1), 93--116.
	
	\bibitem{biecek2021explanatory}
	Biecek, P., Burzykowski, T. (2021).
	\textit{Explanatory Model Analysis}.
	Chapman and Hall/CRC.
	
	\bibitem{cefis2025new}
	Cefis, M., Metulini, R., Carpita, M. (2025).
	A New Dataset for Exploring Actions of Italian Football Matches.
	In \textit{IES 2025}, pp.~624--630.




	
\end{thebibliography}

\end{document}